\documentstyle[11pt,aaspp4]{article}

\def\etal{{\it et~al.\/\ }}
\def\etals{{\it et~al.'s\/\ }}

\def\teff{{\it T$_{\rm eff}$}}
\def\logg{{\rm log~g}}

\def\kms{{km~s$^{-1}$}}

\begin{document}

\title{Boron in the Small Magellanic Cloud
\footnote{Based on observations made with the NASA/ESA
        {\it Hubble Space Telescope}, obtained at the Space Telescope
        Science Institute, which is operated by the Association of
        Universities for Research in Astronomy, Inc., under NASA
        contract NAS 5-26555. These observations are associated
        with proposal GO\#07400.} 
       : A Novel Test of Light Element Production}

\author{A.\,M.\,Brooks\altaffilmark{2} and K.\,A.\,Venn\altaffilmark{3}}
\affil{Macalester College, Saint Paul, MN, 55105}

\author{D.\,L.\,Lambert}
\affil{Univ.\,Texas Austin, Austin, TX, 78712}

\author{M.\,Lemke}
\affil{Dr.\,Karl Remeis Sternwarte, Bamberg, Germany}

\author{K.\,Cunha\altaffilmark{4}}
\affil{Observatorio Nacional - CNPq, Rio de Janeiro, Brazil}

\and 

\author{V.V.\,Smith}
\affil{Univ.\,Texas El Paso, El Paso, TX, 79968}

\altaffiltext{2}
{Columbia University, Department of Astronomy, 
New York, NY, 10027}
\altaffiltext{3}
{University of Minnesota, Department of Astronomy, 
Minneapolis, MN, 55455}
\altaffiltext{4}
{Univ.\,Texas El Paso, El Paso, TX, 79968}

\begin{abstract}

Hubble Space Telescope STIS observations of the \ion{B}{3} resonance 
line at 2066~\AA\ have been obtained and analyzed for two Small
Magellanic Cloud (SMC) B-type stars.  While boron is not detected in
either star, upper limits to the boron abundance are set, with
12+log(B/H) $\le$ 1.6 for both AV\,304 and NGC\,346-637.
The upper limits are consistent with the relationship between boron and oxygen 
previously reported for Galactic disk stars.   The SMC upper limits are
discussed in light of that galaxy's star formation history, 
present oxygen abundance, and its present cosmic ray flux.
 
The UV spectrum has also been used to determine the iron-group 
abundances in the SMC stars.  For AV\,304, [Fe/H]=$-$0.6$\pm$0.2, 
from both an absolute and a differential analysis (with respect to 
the Galactic B-type star HD\,36591).  This is consistent with results 
from A-F supergiants in the SMC.     A lower iron abundance
is found for NGC\,346-637, [Fe/H]=$-$1.0$\pm$0.3, but this is 
in good agreement with the supergiant iron abundances in NGC\,330,
another young SMC cluster.  We propose NGC\,346-637 may be an
unrecognized binary though, which complicates its spectral analysis. 

\end{abstract}


\section{Introduction \label{intro}}

Most of the elements in the Universe have been created by either
Big Bang nucleosynthesis or stellar nucleosynthesis.
There are a few elements, however, that owe their existence, in part
or in whole, to another site and other nuclear processes; that is, to
spallation (and fusion) reactions involving galactic cosmic rays and
ambient interstellar nuclei.    Among this minority are lithium,
beryllium, and boron (LiBeB).   Understanding this trio's origins
requires extensive observations of their abundances in a variety
of objects.
Detailed understanding of boron's nucleosynthesis has lagged behind
that of the two other light elements in large part because lithium
and beryllium are observable in optical spectra, but detection of
boron demands UV spectroscopy.   An extensive body of observational
data on abundances of lithium and beryllium is available for 
theoretical considerations (c.f., Boesgaard \etal 2001, 
Deliyannis \etal 1998, Smith \etal 1998, and references therein), 
but little is presently known about boron.

Reeves, Fowler, \& Hoyle (1970; with refinements by 
Meneguzzi, Audouze, \& Reeves 1971) proposed that LiBeB are 
produced by spallation reactions 
(e.g., $p$ + O $\rightarrow$ Li, Be, or B) 
and a fusion reaction ($\alpha$ + $\alpha$ $\rightarrow$ $^6$Li and $^7$Li)
between a cosmic ray and an ambient nucleus in interstellar gas. 
In the local Galactic interstellar medium, the dominant contribution
to the synthesis of the boron isotopes $^{10}$B and $^{11}$B is thought
to be cosmic ray protons spallating interstellar $^{16}$O nuclei.
An alternative source of $^{11}$B, but not
$^{10}$B has been proposed by Woosley \etal (1990); 
neutrino-induced spallation of $^{12}$C in the carbon-burning 
shell of a massive star undergoing a Type~II supernovae (the $\nu$-process).

The relative contributions of galactic cosmic rays versus  
supernovae to boron's synthesis are ill-determined at present.
A calibration of these contributions is, in principle, possible
by examining the B/Be ratio, as beryllium production is not 
predicted by the the $\nu$-process.  Analysis of 
B/Be ratios are
close to that predicted from spallation over a wide range of
metallicities  (Duncan \etal 1997, Garcia-Lopez \etal 1998).
However, the uncertainties in B/Be ratios are such that a
contribution ($\le$ 30\%, e.g., Lemoine \etal 1998, Vangioni-Flam \etal 1996) 
to $^{11}$B from Type~II supernovae is not excluded.
Furthermore, cosmic ray spallation theory predicts the ratio
$^{11}$B/$^{10}$B = 2.5, and yet Zhai \& Shaw (1994) measured
a higher ratio, $^{11}$B/$^{10}$B = 4.05, in meteorites.
This result also suggests an additional contribution to the 
$^{11}$B abundance, but it is not proof of boron synthesis in
supernovae.  This is because the isotopic ratio may also be increased
by crafting the energy spectrum of the cosmic ray flux at the
threshold energies for boron production by spallation 
(e.g., Meneguzzi \etal 1971, Prantzos \etal 1993, Lemoine \etal 1998). 

To advance our understanding of the synthesis of boron, it is valuable 
to measure boron abundances in a diverse set of environments.
Here, we describe an attempt to measure the boron abundance in the
Small Magellanic Cloud, where the history of star formation, present
metal abundances, and present cosmic ray flux differ from the local Galactic
values.   For example, in the SMC, oxygen has been well determined 
from both nebulae and stars, and is known to be about 1/4 as abundant 
as in the solar neighborhood (c.f., Korn \etal 2000, Hill 1999,
Venn 1999, and references therein).   
Furthermore, Sreekumar \etal (1993) found that the 
cosmic ray flux in the SMC is no more than 1/5 that 
near the Sun based on EGRET observations that failed 
to detect $\gamma$-rays at energies $\geq$100\,MeV 
($\gamma$-rays are the main decay product of $\pi^0$'s produced 
in the collision of cosmic rays with interstellar atoms). 

Herein, we present boron abundances from the \ion{B}{3} 2066\,\AA\
line from {\it HST} STIS spectra of two SMC main-sequence B-type stars,
AV\,304 and NGC\,346-637.   Both of these stars have been well-studied 
in the optical by Rolleston \etal (1993, 2002), so that atmospheric 
parameters and some elemental abundances are available.    These stars 
show no signs of internal mixing (usually demonstrated through 
enrichments of surface nitrogen abundances), and Rolleston \etal have
found oxygen abundances that are in excellent 
agreement with other SMC oxygen results.  
We have also used the {\it HST} STIS spectra to determine 
the iron-group abundances in the two SMC B-type stars, for
comparison with analyses of cool SMC supergiants.

\section{Target Selection and Observations}

Two SMC	B-type stars, AV\,304 and NGC\,346-637, were selected for 
{\it HST} STIS spectroscopy near the \ion{B}{3} 2066\AA\ line.
These SMC stars are known sharp-lined objects (i.e., low $v$sin$i$), which is a
necessary property to be able to resolve the \ion{B}{3} line in 
the crowded UV spectrum.
%
%
Two main sequence B-type stars in the Galaxy (HD\,36591 and HD\,34078), 
with temperatures similar to the SMC targets (see Table~\ref{atms}), 
were also selected as standards for differential analyses.

The SMC observations (see Table~\ref{obs}) were made with the G230M 
grating (R=30,000) and a 52x0.05 slit to obtain a dispersion of 
0.19~\AA/pixel, or 28 km/s per resolution element.  
The Galactic data are from Venn \etal (2002 = V+02).
Spectra were reduced using the STIS pipeline. 
The spectra were rectified 
using a low-order Legendre polynomial, and offset from vacuum
wavelengths (observed) to air wavelengths (line list, discussed below).  
Additionally, the Galactic spectra were smoothed (3-pixel boxcar smoothing).
The final signal-to-noise of each spectrum is listed in Table~\ref{obs}.
The spectra are shown in Figures~\ref{spec1}, \ref{spec2}, \ref{spec3},
and \ref{spec4}.

\section{The Abundance Analyses}

\subsection{Line List}

The line list and  atomic data originated from the Kurucz 
(1988; CD-18) line list, including all lines in the the 
iron-group, light elements, and heavy elements lists, up to barium, 
and through the third ionization state.
This line list was updated by including the new wavelengths for eight 
\ion{Fe}{3} lines reported by Proffitt \etal (1999) from FTS laboratory 
measurements.   We also updated the atomic data of 172 \ion{Fe}{3} 
lines listed in Kurucz's line list by adopting the oscillator strengths 
and wavelengths from Ekberg (1993).
This is similar to the line list used by V+02,
but over a larger wavelength region (2044 to 2145\AA).
 
Atomic data for the \ion{B}{3} 2${s^2}$S $-$ 2${p^2}$P resonance 
doublet with lines at 2065.8~\AA\ and 2067.3~\AA\ are taken from 
Proffitt \etal (1999; and discussed further by V+02).
The weaker \ion{B}{3} line at 2067.3\,\AA\ is blended with a strong
\ion{Fe}{3} line and weaker \ion{Mn}{3} line, and is not suitable
for boron abundance determinations. 
For all syntheses, an isotopic ratio $^{11}$B/$^{10}$B= 4.0 is 
assumed, the solar system ratio (Zhai \& Shaw 1994, Shima 1963).
This is consistent with the estimates given by Proffitt \etal (1999) 
from their line profile analyses of two sharp-lined B-type stars.  
Uncertainties in the use of this ratio are discussed below.  

\begin{figure}
\epsscale{.8}
\plotone{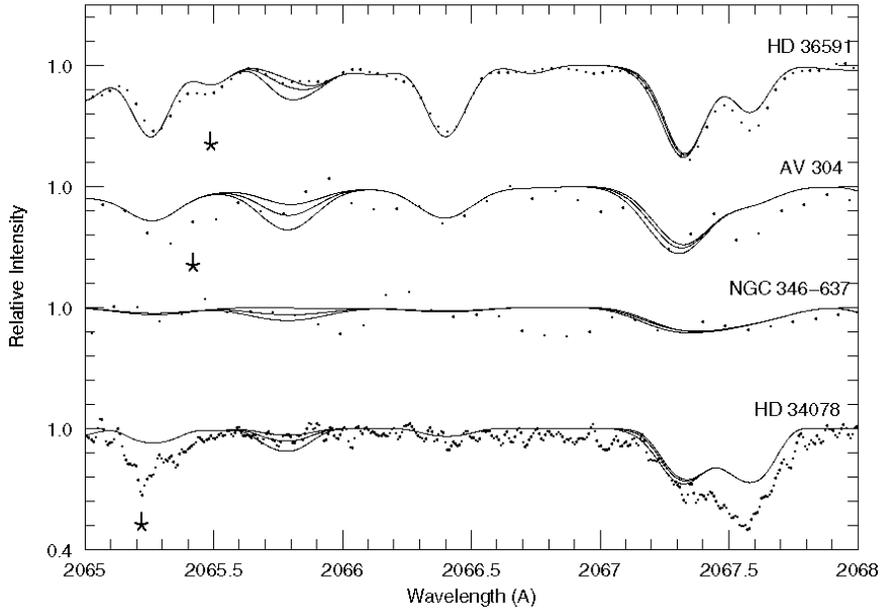}  
\caption{Boron syntheses for all program stars.   
The best fit syntheses are
shown, as well as $\Delta$log(B/H)=$\pm$0.3 for comparison, except for 
NGC\,346-637 where the best fit, 
+0.3, and no boron (12+log(B/H)=$-$10) are shown. 
Interstellar lines are again noted.  The spectra of HD\,36591 and 
HD\,34078 are shown in Venn \etal (2001) for the same wavelength region, 
but the spectra here are synthesized at different temperatures 
(HD\,36591 has been lowered 3.4\% and HD\,34078 has been raised to 
33,000~K, see text).
\label{aly1}}
\end{figure}

Contamination of the spectra by interstellar (IS) lines is considered. 
Proffitt \& Quigley (2001) first noted the importance of IS lines in 
this wavelength range, particularly one \ion{Cr}{2} interstellar line that can 
come close to the \ion{B}{3} $\lambda$2065.8 feature depending on the 
stellar radial velocity.    
Data from Morton (1991) have been used to pinpoint the location of
possible interstellar lines in the spectra and we note those that 
appear in all of our spectra.  The SMC spectra 
contain doubles of each IS line, owing to more than one cloud along the
line-of-sight to these stars (presumably a Galactic component and an 
SMC component).  We have identified four IS lines;
\ion{Cr}{2} (2055.60, 2061.58, and 2065.50) and \ion{Zn}{2} (2062.01).
The locations of the IS lines are noted in the spectrum figures
(see Fig.s~\ref{spec1} through \ref{aly1}).   
They have different locations in each figure because the 
spectra are shown in the stellar rest frames.

The final line list includes 6685 features between 2044 and 2145~\AA. 
All were included in the syntheses but many are negligible contributors.
A few final fine adjustments were made to the line list; 
after an examination of a preliminary syntheses of HD\,36591, 
slight wavelength shifts were made to isolated strong Fe-group 
lines to improve the line synthesis.
These fine adjustments are reported in Table~\ref{offset-lines}.
In addition, after the initial iron-group abundance determination 
was made for HD\,36591 from this line
list, fine adjustments were made to the oscillator strengths of
five lines not previously included due to their grossly
inconsistent fits in comparison to the remaining 64 lines, which 
are also listed in Table~\ref{offset-lines}.  
These five lines were then included in a differential analysis.  
The final line list does a remarkably good job at fitting the 
stellar spectra over this UV range in HD\,36591, AV\,304, and 
HD\,34078.  This is noteworthy because UV line lists are notoriously 
incomplete and/or uncertain in their atomic data.   
Examination of the spectrum figures finds very few missing lines, 
and quite good fits suggesting that the energy 
levels and transition probabilities are fairly accurate.

\subsection{Synthesizing the Spectra \label{syntheses}}

Elemental abundances have been determined from LTE spectral syntheses 
and ATLAS9 model atmospheres (Kurucz 1979, 1988).  The stellar \teff, 
gravity, and projected rotational velocity ($v$sin$i$) values were 
adopted from the literature, see Table~\ref{atms}.  
The \teff\ listed for HD\,36591 and HD\,34078 have been scaled 
down by 3.4\% from that listed in Gies \& Lambert (1992: hereafter GL92) in 
order to remove the increase they applied to their photometic results.  
Corresponding to this lower \teff, the GL92 NLTE nitrogen and oxygen 
abundances have been adjusted by the $\Delta$ values in their Table 9.  
In addition, the GL92 NLTE abundances are based on calculations made 
with Gold (1984) model atmospheres, instead of the more heavily 
line-blanketed Kurucz models.   Thus, we have also applied a correction 
to account for the Gold-Kurucz offsets, 
as tabulated by Cunha \& Lambert (1994, their Table 10).

Other parameters were determined from the syntheses, 
i.e. microturbulence ($\xi$) and radial velocity (as described 
by V+02) and are listed in Table~\ref{abus}.
Macroturbulence ($\xi_{Ma}$) was initially set to the instrumental
broadening values (28~\kms\ for the SMC stars, 15~\kms\ for HD\,36591, 
and 4~\kms\ for HD\,34078).  These values were increased for the 
Galactic stars by 3~\kms\ to best fit the smoothed spectral line
profiles.
ATLAS9 models with 1/10 solar metallicity were used 
for the SMC stars, while solar metallicity models were 
adopted for the two Galactic stars.  
Spectral syntheses were made using the program 
LINFOR\footnote{LINFOR was originally developed by H.\,Holweger,
W.\,Steffen, and W.\,Steenbock at Kiel University.   It has
been upgraded and maintained by M.\,Lemke, with additional
modifications by N.\,Przybilla.}.

\begin{figure}
\epsscale{.8}
\plotone{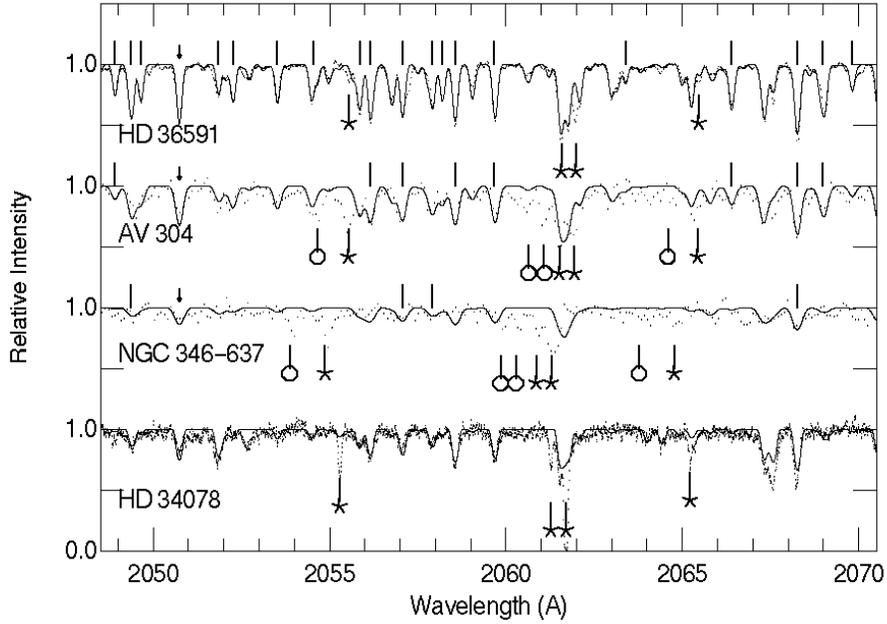}
\caption{Coadded HST STIS spectra for SMC (AV\,304 and NGC\,346-637) and 
Galactic (HD\,36591 and HD\,34078) stars ({\it dotted line})
and their spectrum syntheses ({\it thin line}).   The iron-group 
metallicities in Table~\ref{metals} were used for each synthesis,
i.e., [$M$/H]=0.07, $-$0.6, and $-$1.0 for HD\,36591, AV\,304, and 
NGC\,346-637, respectively.   Solar iron-group abundances are adopted
for HD\,34078 (see text). 
All features listed in Table~\ref{metals} are 
identified by a line {\it above} the feature; features used solely 
for a differential analysis are marked by an arrow.  
Interstellar lines are marked {\it beneath} the 
spectra.  Note that the SMC stars have two sets of IS lines (represented 
by different symbols).  
The Galactic spectra were smoothed for a 3 pixel resolution element. 
\label{spec1}}
\end{figure}

\begin{figure}
\epsscale{.8}
\plotone{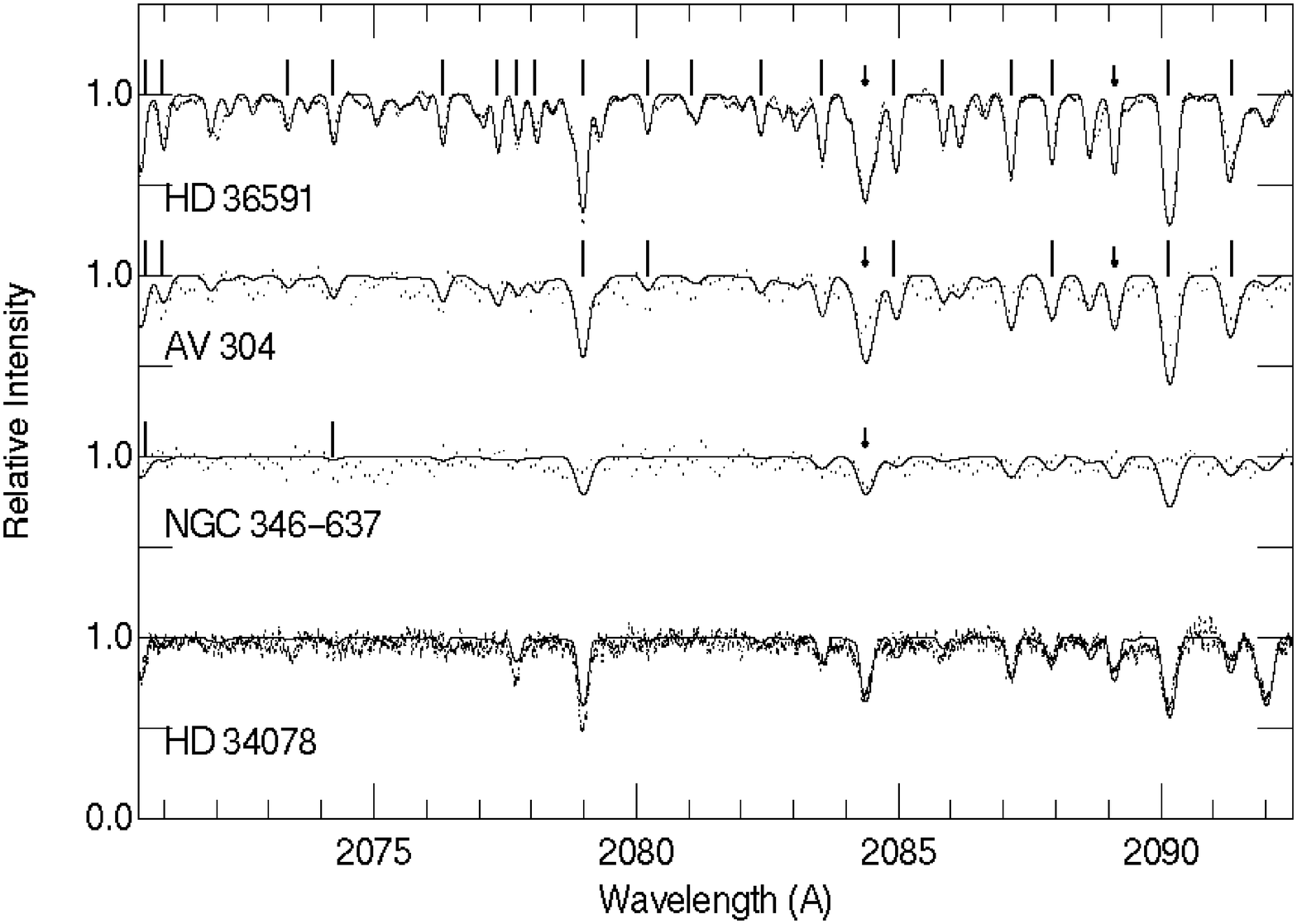}
\caption{See comments in Figure~\ref{spec1}.
\label{spec2}}
\end{figure}

\begin{figure}
\epsscale{.8}
\plotone{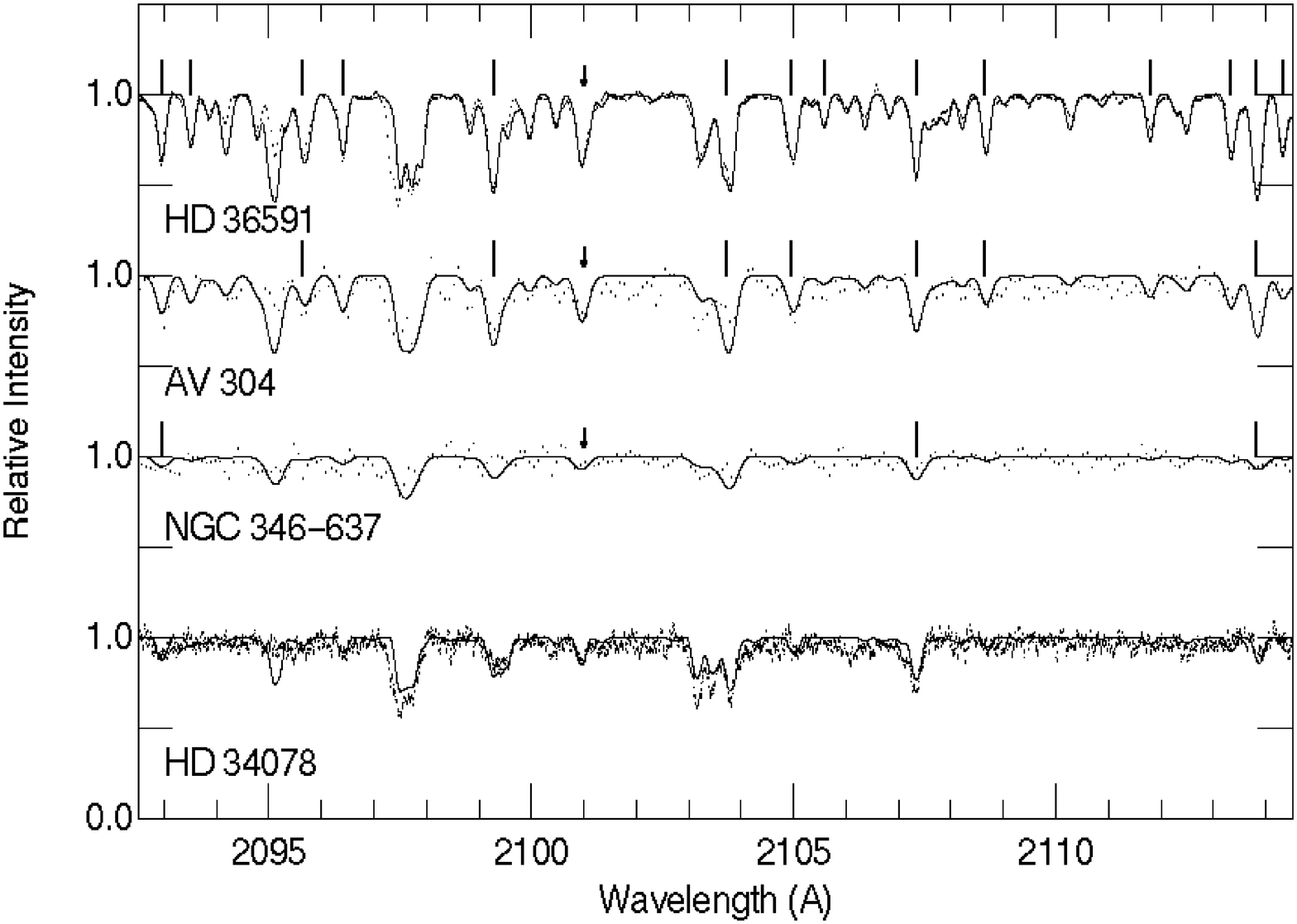}
\caption{See comments in Figure~\ref{spec1}.
\label{spec3}}
\end{figure}

\begin{figure}
\epsscale{.8}
\plotone{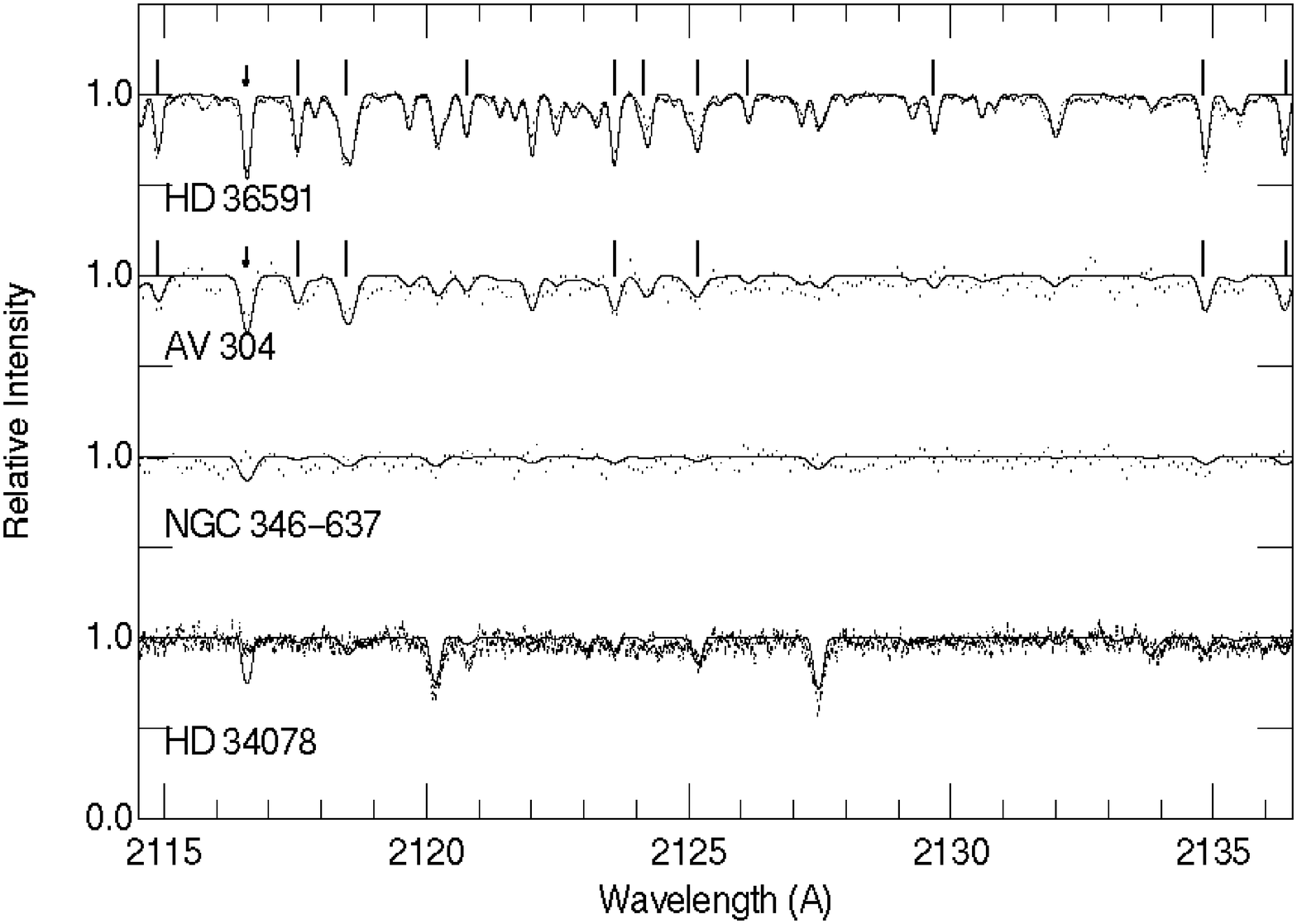}   
\caption{See comments in Figure~\ref{spec1}.
\label{spec4}}
\end{figure}

For the two cooler stars, HD\,36591 (Galactic) and AV\,304 (SMC), 
the spectrum syntheses fit the observed spectra well.   For 
the two hotter stars, HD\,34078 (Galactic) and NGC\,346-637 (SMC),
the initial spectrum syntheses proved to be unsatisfactory.
For HD\,34078, a slightly lower $v$sin$i$ was required to fit its 
sharp line UV spectrum.  Also, the preliminary iron abundance result was 
{\it very} low for a solar neighborhood object; [Fe/H]= $-$1.0.
Since HD\,34078 is one of the hottest stars analysed, then the 3.4\% 
reduction to the GL92 temperature (discussed above) significantly 
affects the iron abundance\footnote
{Note that this iron-group abundance result is also lower than that reported by 
V+02 for HD\,34078 for two reasons.  Firstly, V+02  
synthesized fewer features (twelve in 22~\AA, whereas 25 features are 
synthesized over 70~\AA\ in this paper).  Secondly, the iron abundance 
is {\it very} sensitive to temperature in {\it hot} stars (see the temperature 
sensitivities listed in Table~7 of V+02).   The 3.4\% temperature
effect on abundances was applied to the boron and CNO 
abundances in the discussion by V+02, but not to iron since it was
an insignificant effect for the stars in the discussion in that analysis.}.
Since this star is expected to have a near solar iron abundance, 
as it is a former member of the Orion association, then we found
it is necessary to raise the temperature to 33000~K 
to yield this result.
Accordingly, we have not completed a differential analysis of 
NGC\,346-637 with respect to HD\,34078 because of the uncertainties 
in the parameters of this hot star.   However, HD\,34078 is 
included in Figs~\ref{spec1} through ~\ref{spec4} to show the high  
quality of our line list, even at hot temperatures. 
 
The analysis of NGC\,346-637 also had some difficulties, beginning
with its radial velocity.
We found that a radial velocity of 250~\kms\ was necessary to best fit the 
{\it HST} STIS data, yet Rolleston \etal (1993) reported a value of
$\sim$100~\kms\ for this star.
A large radial velocity is unusual; the velocities for SMC 
stars tends to range from 100 $-$ 180 \kms\ 
(e.g., Venn 1999, Rolleston \etal 1993, Grebel \etal 1996).
In addition, the spectrum synthesis proved to be quite difficult,
with very few clean iron-group lines available to constrain the 
fitting parameters (see Figs.~\ref{spec1} to \ref{spec4}). 
An SMC-like iron-group abundance (discussed below) was found, 
[Fe/H]~=~$-$1.0~$\pm$0.3 (see Table~\ref{metals}),
but our spectrum fits were not very satisfactory.
We examined other spectrum syntheses over a range in atmospheric 
parameters (i.e., \teff, gravity, radial velocity, and $v$sin$i$),
but no significant improvements were found.
We suspect that this star may be an unresolved binary. 
Binarity could explain the large and, apparently, variable radial 
velocity for this star, and could affect the spectrum synthesis
if the companion contributes continuum to the UV spectrum.

\subsection{Iron-group \& Synthesis Parameters}

The iron-group abundances were determined from synthesis 
of specific features in the spectra and results from the 
individual features were averaged.  
In Table~\ref{metals}, the line abundances are listed relative to
the meteoritic abundances from Grevesse \& Sauval (1998), 
e.g. log(Fe)= 7.50 and log(Mn)= 5.53.
The mean abundances in Table~\ref{metals} were calculated
by excluding line abundances that fall more than 2$\sigma$ 
from the mean (of the line-to-line scatter).

We estimate an uncertainty of approximately $\pm$2~\kms\ in
the macroturbulence, based on line profile fitting, 
but only $\pm$1~\kms\ in microturbulence based on line strengths.
Iron-group uncertainties for HD\,36591 and HD\,34078 were previously 
determined by V+02.   These uncertainties are adopted
here for the SMC stars since the atmospheric parameters and analysis 
method are similar. 
NLTE effects are neglected throughout this iron-group
analysis.  Iron-group abundances are determined primarily 
from lines of the dominant ionization species of the elements, 
i.e., \ion{Fe}{3}.

As expected, near solar mean iron-group abundances are found for
HD\,36951, [Fe/H]=$-$0.07 $\pm$0.10 from 46 features.
The iron-group abundance for AV\,304 is [Fe/H]=$-$0.6 $\pm$0.2 from 
24 features, which is identical to a differential line-by-line 
comparison with HD\,36591 (including the five additional features 
with altered oscillator strengths discussed above
for a total of 29 lines). 
This metallicity is similar to Rolleston \etals (1993) results for 
AV\,304's light elements (e.g. Si, Mg), as well as a pioneering 
attempt to determine iron in AV\,304 based on GHRS spectra
by Peters \& Grigsby (1999). 
It is also in good agreement with iron abundances in cooler 
supergiants ($-$1.0 $<$ [Fe/H] $<$ $-$0.5; 
Venn 1999, Hill 1999, 1997, Luck \etal 1998, Russell \& Bessell 1989). 

The mean iron-group abundance for NGC\,346-637 is 
[Fe/H]= $-$1.0 $\pm$0.3, determined from nine features.  
A differential analysis 
of these nine features, plus three more with corrected oscillator 
strengths, with respect to the cooler Galactic star HD\,36591 
yields $-$1.0 $\pm$0.4~dex (recall, due to uncertainties in \teff\
for HD\,34078, we do not present a differential analysis of NGC\,346-637
with this star even though they are both hotter stars).  
The iron-group abundance for NGC\,346-637 is somewhat
lower than that for AV\,304, which could be explained by dilution of
the UV continuum by a companion if NGC\,346-637 is an unrecognized
binary (discussed above). 
However, NGC\,346-637 has an iron abundance that is within the range 
of the results from young supergiants, and it is in excellent 
agreement with the iron abundance found for supergiants in the 
young SMC cluster NGC\,330 (Hill 1999, Luck \etal 1998).

\subsection {Boron Abundances}

LTE boron abundances for the SMC stars are listed in Table~\ref{abus}
following the procedures detailed by Venn \etal (2002).
%
The spectrum synthesis for HD\,36591 is slightly different from that
in V+02 because \teff\ has been lowered 3.4\% from the 
GL92 here; also, the temperature for HD\,34078 is higher, 33,000~K,
to yield solar-like abundances (discussed above).   

In general, no attempt was made to constrain the weak \ion{Mn}{3}~$\lambda$2065.9 
line abundance $a~priori$, which is blended with the 
\ion{B}{3}~$\lambda$2065.8 feature, as described by V+02. 
%
%
However, in the case of AV\,304, an unfortunate noise spike redward of the 
boron feature causes the \ion{Mn}{3} abundance to be indeterminate.  
Therefore, we have adopted the underabundance suggested by the 
iron-group analysis.  In the analysis by V+02, it was 
found that the \ion{Mn}{3} 2065.9 line abundance was generally in 
good agreement with the iron-group determination, with $-$0.3 $\geq$ 
[Mn/Fe] $\le$ $-$0.05.  If [Mn/Fe] = $-$0.3, it does not alter the
boron upper limit for AV\,304.  The iron-group abundance was also 
adopted for the \ion{Mn}{3} 2065.9\AA\ line in NGC\,346-637 and 
HD\,34078 as this feature seems to be insignificant at these hotter 
temperatures (i.e. lowering the 
\ion{Mn}{3} abundance further does not alter our boron upper limits). 

As noted above, an isotopic ratio of $^{11}$B/$^{10}$B = 4.0 is adopted.  
Of course, this ratio is undetermined in the 
SMC, and a purely GCR spallation model predicts $^{11}$B/$^{10}$B = 2.5.
Calculations show that the difference in the boron abundance yielded by 
using a ratio 
of 2.5 or 4.0 is negligible, $\Delta$log(B/H) $\le$ 0.05 for both SMC stars.

To compute the boron abundance {\it uncertainties} 
independent of uncertainties in 
\ion{Mn}{3} $\lambda$2065.9, it was found necessary to fix the 
\ion{Mn}{3} line abundance {\it a priori}.  The \ion{Mn}{3} line 
abundance was set to the best fit value (in Table~\ref{abus}).
A second method of setting the \ion{Mn}{3} abundance according 
to the Fe-group uncertainties for each parameter was examined 
by V+02 and found to yield similar boron uncertainties. 
Table~\ref{boron-unc} shows that the most significant uncertainty 
in the boron abundances tends to be the continuum placement 
(thus, S/N of the data).
The two hottest stars, HD\,34078 and NGC\,346-637, are also sensitive
to the atmospheric
parameters, particularly \teff.  In general, hot stars are 
not the best boron indicators. 
While the result for NGC\,346-637 is consistent with a lower boron 
abundance in the SMC, it is possible that {\it no} boron exists in 
NGC\,346-637 
to within the errors caused by the low signal-to-noise spectrum.  
This is shown in   
Fig.~\ref{aly1}, where the best fit upper limit is shown, with +0.3 and 
12+log(B/H) = $-$10 for comparison.  
Finally, in Table~\ref{abus}, the LTE boron abundances are corrected 
for NLTE effects using calculations reported in V+02.

\section{Discussion }

\subsection{Boron and Oxygen in the SMC and the Galaxy \label{initials}}

Boron synthesis, whether controlled by spallation in the interstellar
medium or by Type II supernovae, is coupled to the growth of the
oxygen abundance. Therefore, it is of interest to establish the 
relationship between boron and oxygen in the SMC.

At present, our boron upper limits in two young stars are the 
sole data points on boron (12 + $\log$(B/H) $\leq$ 1.6 in both).
The oxygen abundances in our SMC targets are listed in 
Table~\ref{atms}, and are in excellent agreement with oxygen
results from other young SMC stars and H\,{\sc ii} regions. 
Oxygen abundances in the SMC have been determined from
H\,{\sc ii} regions, hot and cool supergiants, and B-type
main sequence stars. The results are pleasingly consistent -
see summary by Venn (1999, Table~9). More recent results include
additional analyses of B-type stars by Korn \etal (2000)
and Dufton \etal (2000), who found mean abundances 12 + $\log$(O/H)
= 8.15 and 8.0, respectively. 

\begin{figure}
\epsscale{.8}
\plotone{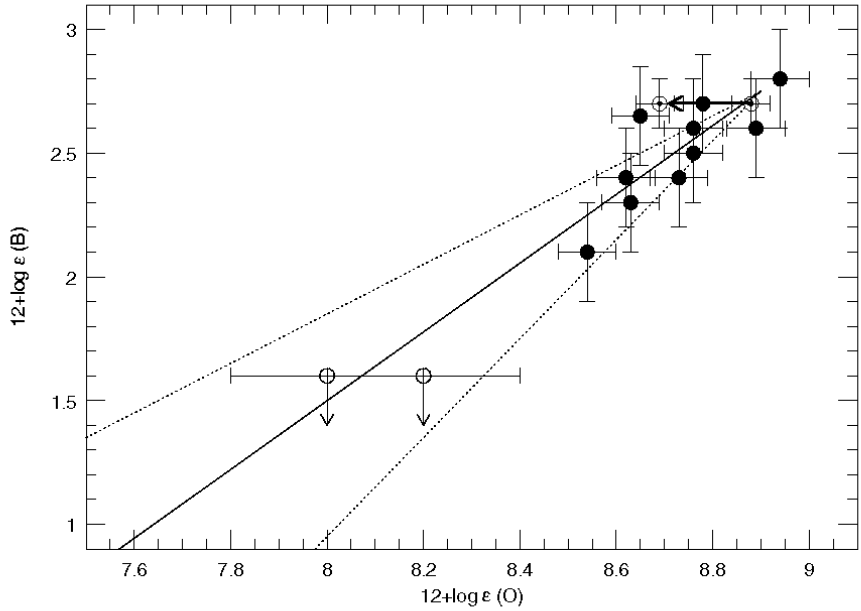}  
\caption{Galactic disk stars' boron and oxygen relationship from
Smith \etal (2001; {\it solid circles}).   
In this log-log plot of abundances, the 
relationship has slope of 1.39 $\pm$0.08 ({\it solid line}), or
N(B) is proportional to N(O)$^{1.4}$.    A new solar abundance by
Allende-Prieto, Lambert, \& Asplund (2001) of 12+log(O/H)=8.69 $\pm$0.05
is noted, and would change the slope slightly to 1.5.
Also shown are the predictions for a linear (slope of 1.0) 
and quadratic (steeper slope of 2.0) relationship between boron 
and oxygen.    We note that the two SMC B-star upper-limit abundances
({\it open circles}) are in good agreement with the Galactic disk slope.
However, the linear relationship cannot be ruled out if boron has been
depleted by rotational mixing. 
\label{boplot}}
\end{figure}

A comparison of boron and oxygen can also be made for Galactic stars
and the local interstellar medium.    In Table~\ref{galb}, we list boron
abundances in local Pop.~I stars and nebulae, and adopt 12 + log(B/H) = 2.5
in the solar neighborhood.  Furthermore, we adopt 12 + $\log$(O/H) = 8.7 
(see Allende Prieto \etal 2001) for the present local oxygen abundance.
Additional local boron abundances have been determined 
over a wide range in metallicities, and several studies have reported 
on the boron-oxygen correlation (e.g., Garcia-Lopez \etal 1998,
Duncan \etal 1997).
Smith, Cunha, \& King (2001) provide a reassessment of the 
relation between boron and oxygen in Galactic F- and G-type dwarfs.
The oxygen abundances in stars
where the B\,{\sc i} 2497 \AA\ line was observed led
to a correlation represented by B/H $\propto$ (O/H)$^m$
with $m = 1.4\pm0.1$
for stars with $-$0.4 $<$ [O/H] $<$ +0.2, see Fig.~\ref{boplot}. 
(Note that our adopted initial local Galactic abundances for boron
and oxygen lay on this line in Fig.~\ref{boplot}).
On including results from the literature to extend the relation to 
lower metallicities, Smith \etal\ found $m$ $\simeq$ 1.0 or 1.4 for
stars with [Fe/H] $\leq -1.0$ depending on the adopted set of oxygen 
abundances in the metal-poor stars.   The uncertainties in oxygen
in metal-poor stars is an ongoing debate at present (c.f., Lambert 2001).
Also, it should be noted that the abundances of beryllium, 
a product solely of spallation, behave in a very similar way to boron,
and the B/Be ratio is approximately constant from
solar metallicity stars to the most metal-poor (Boesgaard \etal 1999). 

In Fig.~\ref{boplot}, we show the oxygen abundances and boron 
upper limits established for the two SMC stars reported here.
The solid line is the Smith \etal fit to the B--O relation for 
the Galactic stars with B $\alpha$ O$^{1.4}$, while the dashed 
lines indicate linear and quadratic trends of B with O.  
If taken at face value,
the results for the two SMC stars would suggest that the
boron-oxygen relationship in the SMC is in excellent agreement with 
that of the Galactic disk.

\subsection{Boron Depletion Mechanisms}

A discussion of boron synthesis in the SMC assumes that 
boron has not been depleted in the stellar atmosphere,
and should represent the present day abundance in the interstellar 
medium.  Observations of Galactic B-type stars, however, show that
boron depletion is not uncommon 
(Venn \etal 2002, Proffitt \& Quigley 2001).
Therefore, the possibility of loss of atmospheric boron must
be recognized. 

Several processes are capable of reducing the
boron abundance while a B-type star is on the main sequence.
The most likely process is rotationally-induced mixing, although
mass loss, or mass transfer from an evolved companion, would also
produce the same result.
Mass loss from {\it main sequence} B-type stars is not expected to have 
a significant effect on the surface B abundance (see the discussion in
Fliegner \etal 1996, and V+02).  It would require a mass loss 
rate an order of magnitude larger than observed for Galactic B-type stars 
(Cassinelli \etal 1994).   For SMC stars, we expect the winds to be 
even weaker since winds of hot stars are driven by photon momentum transfer 
through metal line absorption, thus a function of metallicity 
(c.f., Kudritzki \& Puls 2000).

Mass transfer in a close binary system will not affect boron 
in the mass-receiving (i.e., the observed) star without a considerable 
enrichment of nitrogen (c.f., Wellstein 2000).   For AV~304, 
Rolleston \etal (2002) find the nitrogen abundance 
12 + $\log$(N/H) = 6.7 $\pm$ 0.2 from N\,{\sc ii} lines
measured from a VLT UVES spectrum. 
The fact that this abundance is in very good
agreement with those measured in SMC H\,{\sc ii} regions
(Dufour 1984; Russell \& Dopita 1990; Kurt \etal 1999), and  
for other B-type stars (e.g., Korn \etal 2000) eliminates
the idea that mass transfer could have affected its boron abundance.
For NGC 346-637, the picture is unclear. 
Rolleston \etal (1993) could determine only an upper limit to 
the nitrogen abundance: 12 + $\log$(N/H) $\leq$ 7.2, which does 
not entirely exclude the possibility that mass transfer may have 
affected the surface abundances.    Furthermore, we suspect this
star may be an unrecognized binary (see Section~\ref{syntheses}).


Depletion of boron by rotationally-induced mixing cannot be
ruled out. 
Recent models by Heger \& Langer (2000) follow the evolution of
the angular momentum distribution in intermediate-mass stars from
the pre-main sequence through core collapse, and find that rotational
mixing can affect stellar surface abundances of boron and nitrogen.
For rapidly rotating stars, boron can be depleted and nitrogen 
enriched at the surface.
V+02 have confirmed that these models do indeed fit 
the observed abundances of boron and nitrogen in Galactic B-type stars.
Boron is predicted to be depleted ahead of detectable nitrogen enrichment.
While nitrogen-rich stars are predicted and observed to be boron
depleted (Proffitt \& Quigley 2001, V+02), stars showing
a normal nitrogen abundance may also be boron depleted by as much as 1~dex!
Thus, this is a possibility which we cannot exclude for either AV~304 or
NGC~346-637.  To reduce this possibility additional stars would 
have to be observed to search for stars with a higher boron abundance.
(While lithium and beryllium would be depleted too, these
elements are not observable in B-type stars.)

\subsection{Spallation in the SMC}

In the case of the solar neighborhood, the contribution of spallation to
the synthesis of boron is assessable from measurements of the beryllium
abundance in stars; beryllium owes its origins
exclusively to spallation and the B/Be ratio is predictable with
fair certainty. Unfortunately, there are no determinations of the
Be abundance in SMC gas or stars, and one must predict the production
rate of boron from the ingredients controlling it.

Suppose boron is produced by the process
$p_{CR} + $O$_{ISM} \rightarrow$ B, the production rate $dn_B/dt$
involves three factors: the cosmic ray proton flux ($\phi_{CR}$),
 the O abundance in the
interstellar medium ($n(O)$), and the spallation cross-section
($\sigma(O \rightarrow B)$), that is
\begin{equation}
\frac {dn_B}{dt} = \phi_{CR} n(O) \sigma(O \rightarrow B)
\end{equation}
where the energy dependences of the factors 
are suppressed and an
integral over energy is implicit. To this rate must be added contributions
from C and N as targets, and from $\alpha$-particles as projectiles.

The production rate at present in the SMC would appear to be
much less than that in the local regions of the Galaxy because
(i) the oxygen abundance (n(O)) in the SMC is a factor of 4 lower
than in the local interstellar medium (see Section~\ref{initials}), 
and (ii)  the upper limit on the cosmic ray flux in the SMC is at least
a factor of 5 less than the local flux (Sreekumar \etal 1993). 
Taken together, $\frac {dn_B}{dt}$ for the SMC
could be a factor of 20 less than the local rate. 
But the  boron abundance at a given time and location
is an integral of the production rate; a low rate now 
does not of itself preclude higher rates in the past.
Thus, the history of the cosmic ray flux can also contribute
to predicting the boron abundance.

\subsubsection{History of the Cosmic Ray Flux \label{histcrf}}

The history of the cosmic ray flux (=CRF) must be known in order 
to calculate the evolution of the boron abundance.
The confinement time of Galactic cosmic rays
is short (below), and, hence, the present flux is no guide to the
past flux of cosmic rays. 
Measurements of the abundance of secondary fragments
produced in the interstellar medium by nuclear interactions
between cosmic rays and ambient nuclei have provided estimates 
of the confinement time for Galactic cosmic rays.
 Radioctive $^{10}$Be was the first `clock' to
be proposed (Hayakawa \etal 1958). Now, measurements of $^{10}$Be,
$^{26}$Al, $^{36}$Cl, and $^{54}$Mn all indicate a confinement
time of about 15 Myr (Yanasak \etal 2001). 
Clearly, galactic cosmic rays must be continuously replenished if
spallation is to work its magic.

The confinement time for the SMC is probably shorter.   
The magnetic field lines of the SMC are expected to be highly 
disrupted due to interactions with the Galaxy and the LMC.
Mathewson (1986, 1988) and Murai \& Fujimoto (1980) suggest that a close
encounter between the Magellanic Clouds and the Milky Way occurred about
200 Myr ago, and observations show it is likely that this disturbance
disrupted the magnetic field lines of the SMC (Wayte 1990; Haynes \etal
1991; Ye \& Turtle 1991; Goldman 2000).
As cosmic rays are highly ionized particles, they are generally trapped
by the magnetic field lines within a galaxy; 
thus, disrupted magnetic field lines in the SMC may lead to shorter
lifetimes for cosmic rays in the SMC.

Thus, the present CRF in the SMC is not likely to be the same as in the past.
For example, a large burst of star formation that began 2-4~Gyr ago 
enriched the SMC to its present oxygen abundance 
(e.g., Pagel \& Tautvaisiene 1998, Mighell \etal 1998, 
de Freitas Pacheco \etal 1998, Gardiner 1992),
and it should also have produced a higher CRF.
However, the star formation rate is thought to have declined since that 
time, and hence the CRF could also have declined.  We expect the present 
low flux is a recent phenomenon that should have no influence on the 
boron-oxygen correlation.

\subsubsection{Boron vs Oxygen Relationship}

Although the source of the cosmic rays and the identity of the 
acceleration mechanism are uncertain, it is assumed that the 
CRF is linked to the presence of supernovae. 
Thus, given the short confinement time, 
$\phi_{CR} \propto N_{SN}(t)$ is a plausible approximation.
Since oxygen is a product of Type II supernovae, one
may suppose that $dN_{SN}/dt \propto dn(O)/dt$.
If it is assumed that the constant of proportionality between the
CRF and the number of supernovae  is time independent,
equation (1)  predicts that B/H $\propto$ (O/H)$^m$ with $m \equiv 2$.
This result does not depend on an assumption of a time-independent
$dn_B/dt$; star formation in bursts (as in the SMC) does not necessarily 
invalidate the result.   As long as the mean lifetime of the cosmic rays 
is short relative to the durations of the burst and the CRF is
tied to that of the O-producing massive stars, the production
rate integrated over time will result in the quadratic dependence.

Direct comparison of the boron vs oxygen relationship in the Galaxy and the SMC
will be inappropriate if the efficiency of cosmic ray generation
in supernovae is different for the two galaxies.
Nonetheless, it is of interest to make such a comparison.
In the solar neighborhood, we adopt 12 + $\log$(B/H) = 2.5 
and 12 + $\log$(O/H) = 8.7 (see Section~\ref{initials}).  
Adopting 12 + $\log$(O/H) = 8.1 for the SMC,
the quadratic relation predicts 12 + $\log$(B/H) = 1.3 for the SMC,
a value that is 0.3~dex below our measured upper limits. 
If cosmic rays do escape more easily from the SMC (discussed above), 
then the boron production rate per oxygen atom produced by supernovae 
would be smaller in the SMC.   Recognition of this fact would result in a 
predicted boron abundance that is even lower than 12 + $\log$(B/H) = 1.3.

While our boron upper limits are in fair agreement with the 
quadratic prediction, the quadratic approximation is of 
questionable validity, at least for the solar neighborhood. 
Observations of boron and oxygen abundances in stars for which 
boron is undepleted do $not$ show the quadratic dependence 
(see Section~\ref{initials}): as noted, $m = 1.4$ for
Galactic stars of approximately solar metallicity, and a lower
index (e.g., $m \simeq 1$) may be required to fit the
observations of metal-poor stars (Smith \etal 2001).
If $m = 1.4$ is adopted, the predicted SMC boron abundance
(12 + log(B/H) $\sim$ 1.6) is in excellent agreement with our 
upper limits.   
An index $m = 1$ predicts a boron abundance 0.3~dex greater than 
our upper limits for the SMC stars.   While a difference of
0.3~dex is within 3~$\sigma$ of our boron upper-limits, 
we note that this agreement worsens for our best analysed star, AV\,304.
For AV\,304, 12 + log(O/H) = 8.2, thus a slope of $m = 1$ predicts a 
boron abundance that is 0.4~dex larger
than our upper limit.    This result argues against a purely
linear relation between oxygen and boron in the SMC,
unless we need to consider rotationally induced depletion,
and/or the possibility of less efficient 
production and/or retention of cosmic rays in the SMC. 

Other schemes may also be devised that allow for values 
of $m$ less than 2.  In the context of spallation,
a simple way is to consider the `reverse' of the `direct'
process introduced above.   For the reverse reaction 
O$_{CR}$ + $p_{ISM} \rightarrow$ B, and a boron-oxygen relation
with $m$ = 1, that is B $\propto$ O is expected. 
The reverse rate is often neglected because it 
produces higher velocity boron nuclei than the
direct process, and the losses before the nuclei 
are thermalized are greater than for directly produced boron.
Furthermore, when the interstellar medium has approximately solar abundances
of C, N, and O, the reverse process is much less effective 
than the direct process.  For example, in the early Galaxy, the oxygen
abundance in the 
interstellar medium was lower, but the cosmic rays, if they
were accelerated material from supernova ejecta, may have had
a similar oxygen abundance to contemporary cosmic rays. This circumstance
would favor the reverse rate.  In the limit that the reverse process 
was dominant at low metallicities, one expects 
a switch from $m = 1$ to $m = 2$ as O/H increases in simple galactic 
models. 
And, in fact, depending on the choice for the oxygen abundances 
in metal-poor stars, there is a trend from $m \simeq 1$  at low 
metallicity to a higher value ($m = 1.4$) at solar metallicity 
(see Smith \etal 2001). 

Finally, two other scenarios may be mentioned. 
First, if cosmic rays are not of galactic origin, their flux 
would be independent of the supernova rate, which would affect
all boron-oxygen relations.   However, there is ample evidence 
that the extragalactic component to Galactic cosmic rays is 
small (e.g., Pannuti 2000, Dickel 1974, Butt \etal 2001),
including the fact that the SMC does not have the same CRF as
seen in the local Galaxy (Sreekumar \etal 1993).  
Second, if boron is primarily a product of neutrino-induced
spallation in Type~II supernovae, then one expects $m \simeq 1$.
However, the contribution by this process is currently estimated
at $\le$ 30\% in the Galaxy 
(e.g., Lemoine \etal 1998, Vangioni-Flam \etal 1996).

\section{Conclusions}

We have analysed {\it HST} STIS observations of the B\,{\sc iii} 
resonance line at 2066 \AA\ for two SMC B-type stars. The upper limits
corrected for small non-LTE effects are 12 = $\log$(B/H) $\leq 1.6$ for
both AV 304 and NGC 346-637. Unless the stars have internally depleted
boron by a large factor, we show that the upper limits are
plausibly consistent with the hypothesis that boron is a product of
spallation induced by cosmic rays.   Significant production by 
neutrino-induced spallation of $^{12}$C in Type II supernovae is probably
excluded unless the initial boron abundance was a factor of 2 higher
than our upper limits. 

The UV line list is quite excellent for spectrum synthesis at Galactic
and SMC metallicity in this temperature range.   
For AV\,304, we find [Fe/H]=$-$0.6$\pm$0.2 
from both an absolute and differential analysis with HD\,36591.  
This is consistent with results from the A-F supergiants 
in the SMC.  In comparison, the fewer and weaker iron-group lines 
in the spectrum of NGC\,346-637 result in a less certain abundance,
[Fe/H]=$-$1.0$\pm$0.3.
We also suggest that this star may be an unresolved binary.

\acknowledgments
Support for proposal GO\#08161 was provided by NASA through a
grant from the Space Telescope Science Institute, which is
operated by the Association of Universities for Research in Astronomy, 
Inc., under NASA contract NAS 5-26555. 
KAV and AB acknowledge additional research support from Macalester College 
and the Luce Foundation through a Clare Boothe Luce Professorship award. 
We would like thank Grace Mitchell (STScI) for expertise in reducing 
the STIS data. 


\clearpage
\begin{deluxetable}{lcrrclcl}
\tablecaption{Atmospheric Parameters from the Literature \label{atms}}
\tablewidth{0pt}
\tablehead{
\colhead{Star} & \colhead{SpType} & \colhead{\teff} & \colhead{\logg}  & 
\colhead{ $v$\,sin\,$i$}  &  \colhead{ 12+log($O/H$) } &
\colhead{ 12+log($N/H$) } &
\colhead{REF}   \\[.2ex]
\colhead{} & \colhead{} & \colhead{(K)} & 
\colhead{} & \colhead{(\kms)} & 
\colhead{(NLTE)} & \colhead{(NLTE)} &
\colhead{}  } 
\startdata
HD\,36591  & B1IV & 26449 & 4.15 & 11 & 8.67 $\pm$0.22 & 7.69 $\pm$0.10  & GL92  \nl
------     & B1V  & 26330 & 4.21 & \nodata & 8.54 $\pm$0.05 & 7.64 $\pm$0.05  & CL94  \nl
AV\,304    & B0   & 27500 & 3.80 & 14 & 8.2 $\pm$0.2 & 6.7 $\pm$0.2 & R+02 \nl
HD\,34078  & O9.5V & 33000\tablenotemark{a} & 4.07 & 20\tablenotemark{b}& 8.34 $\pm$0.25 & 7.25 $\pm$0.09 & GL92  \nl
NGC\,346-637 & B0V & 30500 & 4.00 & 28 & 8.0 $\pm$0.2 & $\le$ 7.2 & R+93 \nl
\enddata
\tablecomments{The atmospheric parameters listed here were adopted from ATLAS9 
models and spectral syntheses.  Solar metallicity was used for HD\,36591 and 
HD\,34078, while [Fe/H]=$-$1.0 models were used for SMC stars.  
The parameters from
GL92 were adopted for HD\,36591, with \teff\, lowered by 3.4\%.  The O and N abundances have been adjusted to this 
lower \teff\, scale (see $\Delta$ in Table 9 of GL92), and corrected for the use of Gold NTLE models rather than 
Kurucz (see Venn \etal 2001 for a full discussion).
GL92 = Gies \& Lambert (1992), 
CL94 = Cunha \& Lambert (1994),
R+02 = Rolleston \etal (2002),
R+93 = Rolleston \etal (1993). }
\tablenotetext{a}{Temperature was raised to 33,000 K from the GL92 corrected value of 30352 K, because we
expect that HD\,34078 should have near solar Fe abundances as it is a runaway Orion star.} 
\tablenotetext{b}{While GL92 found $v$\,sin\,$i$ in this star to be 30$\pm$3 \kms, we found that 
a lower $v$\,sin\,$i$ of 20 \kms\ was required to fit the sharp lines in this spectrum.}
\end{deluxetable}
\clearpage

\clearpage
\begin{deluxetable}{lccrrr}
\tablecaption{HST STIS Observing Information for Galactic B-stars \label{obs}}
\tablewidth{0pt}
\tablehead{
\colhead{Star} & \colhead{V} & \colhead{Grat/Slit} & 
\colhead{Exposure(s)} & \colhead{Date} & \colhead{S/N}
} 
\startdata
AV\,304	     & 14.98 & G230M     &  19950 at $\lambda_c$2095 & 26 OCT 99 & 50 \nl
------	     &       & 52x0.05   &  +2460 at $\lambda_c$2095 & 29 SEPT 00 & \nl
------	     &	     &		 &  +22680 at $\lambda_c$2095 & all CVZ   &  \nl
HD\,34078    &  5.96 & E230H     &   432s at $\lambda_c$2063 & 15 MAR 00 & 55 \nl
------       &       & 0.1x0.03  &  +432s at $\lambda_c$2013  \nl
HD\,36591    &  5.34 & E230M     &    432s at $\lambda_c$2124 & 09 FEB 99 & 100 \nl
------       &       & 0.2x0.05ND &  +432s at $\lambda_c$2124  \nl
------       &       &           &   +432s at $\lambda_c$2124  \nl
------       &       &           &   +434s at $\lambda_c$1978  \nl 
------       &       &           &   +434s at $\lambda_c$1978  \nl 
------       &       &           &   +432s at $\lambda_c$2269  \nl 
------       &       &           &   +432s at $\lambda_c$2269  \nl 
NGC\,346-637 & 14.98 & G230M     &    2376s at $\lambda_c$2095 & 25 OCT 99 & 30 \nl
------       &       & 52x0.05   &  +12944s at $\lambda_c$2095 \nl
------       &	     &		 &  +2376s at $\lambda_c$2095 & 2 OCT 00 & \nl
------       &	     &  	 &  +9708s at $\lambda_c$2095  \nl
\enddata
\end{deluxetable}
\clearpage



\begin{deluxetable}{lllr}
\tablewidth{30pc}
\tablecaption{Iron-Group Wavelength Offsets \label{offset-lines}}
\tablehead{
\colhead{Elem} & \colhead{$\lambda$(KUR)} & 
\colhead{ $\lambda$(NEW) } & \colhead{log $gf$}}
\startdata
\ion{Mn}{3} & 2048.949 &   2048.918 & \nodata \\
\ion{Mn}{3} & 2049.357 &   2049.314 & \nodata \\ 
\ion{Mn}{3} & 2049.682 &   2049.663 & \nodata \\
\ion{Mn}{3} & 2063.337 &   2063.397 & \nodata \\
\ion{Mn}{3} & 2065.886 &   2065.892$^a$ & $-$0.241\tablenotemark{^a} \\
\\
\ion{Fe}{3} & 2050.743 &  \nodata  & 0.19    \\
\ion{Fe}{3} & 2052.271 &  \nodata\tablenotemark{b}  & \nodata \\
\ion{Fe}{3} & 2053.524 &  \nodata\tablenotemark{b}  & \nodata  \\
\ion{Fe}{3} & 2054.492 &  \nodata\tablenotemark{b} & \nodata \\
\ion{Fe}{3} & 2055.863 &  2055.859 & \nodata \\
\ion{Fe}{3} & 2056.152 &  2056.156 & \nodata \\
\ion{Fe}{3} & 2057.059 &  2057.072 & \nodata \\
\ion{Fe}{3} & 2057.928 &  2057.925 & \nodata \\
\ion{Fe}{3} & 2058.209 &  2058.205 & \nodata \\
\ion{Fe}{3} & 2058.566 &  \nodata\tablenotemark{b} & \nodata \\
\ion{Fe}{3} & 2064.980 &  \nodata\tablenotemark{b} & \nodata \\
\ion{Fe}{3} & 2068.249 &  2068.263 & \nodata \\
\ion{Fe}{3} & 2070.539 &  2070.561 & \nodata \\
\ion{Fe}{3} & 2070.976 &  2070.996 & \nodata \\
\ion{Fe}{3} & 2076.322 &  2076.318 & \nodata \\
\ion{Fe}{3} & 2080.220 &  \nodata\tablenotemark{b} & \nodata \\
\ion{Fe}{3} & 2082.384 &  2082.377 & \nodata \\
\ion{Fe}{3} & 2084.376 &  \nodata  & 0.99  \\
\ion{Fe}{3} & 2089.093 &  2089.120 & 0.28  \\
\ion{Fe}{3} & 2093.505 &  2093.512 & \nodata \\
\ion{Fe}{3} & 2096.426 &  2096.417 & \nodata \\
\ion{Fe}{3} & 2099.226 &  \nodata\tablenotemark{b} & \nodata \\
\ion{Fe}{3} & 2100.966 &  2100.950 & 0.04  \\
\ion{Fe}{3} & 2107.322 &  2107.339 & \nodata \\
\ion{Fe}{3} & 2108.679 &  2108.684 & \nodata \\
\ion{Fe}{3} & 2116.593 &  2116.583 & 0.24  \\
\enddata
\tablenotetext{^a}{Proffit \etal (1999)}
\tablenotetext{b}{Changed to E93, then shifted back to Kurucz. Kept E93
gf values.}
\end{deluxetable}
\clearpage

\clearpage
\begin{deluxetable}{lccc}
\tablecaption{Boron Abundances from STIS Spectroscopy
              \label{abus}}
\tablewidth{0pt}
\tablehead{
\colhead{Star} & \colhead{\underline{ \ RV \ \ \ \ $\xi_{\rm macro}$ \ \ \ \ $\xi$ \ }} &
\colhead{\underline{ \ \ \ 12+log($B$/H) \ \ \ }} & \colhead{ {\it 12+log(Mn~III)}} \\[.2ex] 
\colhead{} & \colhead{ \ \ (km\,s$^{-1}$) } &   
\colhead{ LTE \ \ \ \ \ \ \ \ NLTE }  &
\colhead{ {\it $\lambda$2065.9}} 
} 
\startdata
HD\,36591  & \ 29 \ \ \ \ \, 18 \ \ \ \ \ 2  & $\le$1.36 \ \ \ \ \ \  $\le$1.27 & {\it 5.31} \nl
AV\,304    & 145 \ \ \ \ \  30  \ \ \ \ \  3   & $\le$1.7   \ \ \ \ \ \ $\le$1.6  & {\it 4.9} \nl
HD\,34078  & \ 56 \ \ \ \ \ \ 7 \ \ \ \ \ \ 3 & $\le$2.4   \ \ \ \ \ \ $\le$2.5  & {\it 5.4}  \nl
NGC\,346-637 & 254 \ \ \ \ \  30   \ \ \ \ \ 2 & $\le$1.5   \ \ \ \ \ \  $\le$1.6  & {\it 4.6}  \nl
\enddata
\tablecomments{Abundances have been determined from spectrum syntheses using the 
model atmosphere parameters listed here and in Table~\ref{atms}.
Radial velocities, $\xi$, and $\xi_{\rm macro}$ values are determined 
from the iron-group 
features.  Radial velocities (RV) have been corrected for vacuum-to-air
wavelength offsets.  NLTE corrections are from calculations in 
Venn \etal (2001). 
The \ion{B}{3} $\lambda$2065.8 
and \ion{Mn}{3} $\lambda$2065.9 abundances were varied together 
in order to achieve the best fit, thus we report the best fit 
\ion{Mn}{3} abundance here in italics.}
\end{deluxetable}
\clearpage

\begin{deluxetable}{lrrrr}
\tablewidth{0pc}
\tablecaption{Iron-group Abundance Results \label{metals}}
\tablehead{
\colhead{$\lambda$} & \colhead{Element(s)} &
\colhead{ [$M$/H] } & \colhead{ [$M$/H] } & 
\colhead{ [$M$/H] } \\[.2ex] 
\colhead{(\AA)} & \colhead {} &
\colhead{36591} & \colhead{AV304} & \colhead{NGC}}
\startdata
2048.92 & \ion{Mn}{3}   & $-$0.09   & $-$0.6      & \nodata 	\\
2049.37 & \ion{Fe}{3}+\ion{Mn}{3} & $-$0.10   & \nodata  & $-$0.4 \\
2049.66 & \ion{Mn}{3}   &   0.00    & \nodata     & \nodata 	\\
2050.74 & \ion{Fe}{3}   &   fixed   & {\it 1.4}   &   0.93  	\\
2051.85 & \ion{Fe}{3}+\ion{Fe}{4} & +0.05 & \nodata & \nodata 	\\
2052.27 & \ion{Fe}{3}   & $-$0.09   & \nodata 	  & \nodata 	\\
2053.52 & \ion{Fe}{3}   & $-$0.24   & \nodata 	  & \nodata 	\\
2054.56 & \ion{Fe}{3}x3 & $-$0.08   & \nodata     & \nodata 	\\
2055.86 & \ion{Fe}{3}   & {\it +0.19} & \nodata   & \nodata 	\\
2056.16 & \ion{Fe}{3}   & +0.08     & $-$0.3      & \nodata 	\\
2057.07 & \ion{Fe}{3}   & $-$0.08   & $-$0.4	  & $-$1.3  	\\
2057.93 & \ion{Fe}{3}   & $-$0.15   & \nodata  	  & $-$1.0  	\\
2058.21 & \ion{Fe}{3}   & $-$0.03   & \nodata     & \nodata 	\\
2058.57 & \ion{Fe}{3}   & {\it +0.23} & $-$0.8    & \nodata     \\
2059.67 & \ion{Fe}{3}   & +0.11     & {\it $-$1.3} & \nodata 	\\
2063.40 & \ion{Mn}{3}   & $-$0.13   & \nodata 	  & \nodata 	\\
2066.40 & \ion{Mn}{3}x2+\ion{Ni}{3} & $-$0.25 & $-$0.5 & \nodata \\
2068.26 & \ion{Fe}{3}   & {\it +0.26} & $-$0.2    & $-$1.4  	\\
2068.99 & \ion{Fe}{3}+\ion{Mn}{3}+\ion{Cr}{3} & $-$0.25 & $-$0.9 & \nodata \\
2069.82 & \ion{Fe}{3}+\ion{Mn}{3} &  0.00  & \nodata  & \nodata \\
2070.56 & \ion{Fe}{3}   & {\it +0.32} & {\it $-$1.2} & $-$1.0  	\\
2070.98 & \ion{Fe}{3}x3 & {\it $-$0.35} & {\it 0.0} & \nodata \\
2073.35 & \ion{Fe}{3}+\ion{Mn}{3} & $-$0.05 & \nodata & \nodata \\
2074.23 & \ion{Fe}{3}   & $-$0.02   & \nodata 	  & {\it +0.4} 	\\
2076.32 & \ion{Fe}{3}   & {\it $-$0.54} & \nodata & \nodata 	\\
2077.36 & \ion{Mn}{3}+\ion{Co}{3} & $-$0.05 & \nodata & \nodata \\
2077.74 & \ion{Fe}{3}+\ion{Fe}{4} & {\it +0.22} & \nodata & \nodata \\
\nl
\tablebreak
\nl
2078.08 & \ion{Fe}{3}+\ion{Mn}{3} & {\it $-$0.35} & \nodata  & \nodata \\
2079.00 & \ion{Fe}{3}x4 & {\it +0.20}   & $-$0.7   	& \nodata \\
2080.22 & \ion{Fe}{3}   & $-$0.07   & {\it +0.2}	& \nodata \\
2081.08 & \ion{Mn}{3}x2+\ion{Co}{3} & $-$0.02   & \nodata & \nodata \\
2082.38 & \ion{Fe}{3}   & $-$0.13   & \nodata 		& \nodata \\
2083.55 & \ion{Fe}{3}   & +0.09     & \nodata 		& \nodata \\
2084.36 & \ion{Fe}{3}x3+\ion{Mn}{3}x2 & fixed   & +1.3  & {\it 2.03} \\ 
2084.93	& \ion{Fe}{3}x2 & $-$0.05   & $-$0.8  		& \nodata \\
2085.84	& \ion{Fe}{3}+\ion{Cr}{3} & +0.08     & \nodata & \nodata \\
2087.15 & \ion{Fe}{3}   & {\it +0.16}   & \nodata 	& \nodata \\
2087.93 & \ion{Fe}{3}   & $-$0.16   & $-$0.6  		& \nodata \\
2089.12 & \ion{Fe}{3}   &  fixed    &   0.9   		& \nodata \\
2090.16 & \ion{Fe}{3}x4+\ion{Mn}{3}x3 &  0.00  & {\it $-$1.3} & \nodata \\
2091.35 & \ion{Fe}{3}x2 & {\it $-$0.40} & $-$1.0	& \nodata \\
2092.97 & \ion{Fe}{3}   &  0.00     & \nodata 		& $-$0.8  \\
2093.51 & \ion{Fe}{3}   & $-$0.08   & \nodata 		& \nodata \\
2095.66 & \ion{Fe}{3}x3 & $-$0.10   & $-$0.3  		& \nodata \\
2096.42 & \ion{Fe}{3}   & +0.13     & \nodata 		& \nodata \\
2099.30	& \ion{Fe}{3}x2 & $-$0.22   & $-$1.0 		& \nodata \\
2101.04	& \ion{Fe}{3}+\ion{Mn}{3} &  fixed    &  0.7    &   0.83  \\
2103.74 & \ion{Fe}{3}x4 & {\it $-$0.40} & {\it $-$1.5}  & \nodata \\
2104.96 & \ion{Fe}{3}x2+\ion{Cr}{3}+\ion{Ni}{3} & +0.02  & $-$0.3  & \nodata \\
2105.59 & \ion{Ni}{3}   & $-$0.13   & \nodata 		& \nodata \\
2107.34 & \ion{Fe}{3}   & +0.11     & $-$0.7  		& {\it $-$1.7} \\
2108.64 & \ion{Fe}{3}+\ion{Mn}{3} & $-$0.15   & $-$0.5  & \nodata \\
\nl
\tablebreak
\nl
2111.80 & \ion{Fe}{3}x2 & {\it $-$0.43} & \nodata  & \nodata 	\\
2113.34 & \ion{Fe}{3}x2+\ion{Mn}{3} & $-$0.25 & \nodata  & \nodata \\
2113.83 & \ion{Fe}{3}x2+\ion{Cr}{3} & $-$0.20  & $-$0.6  & $-$1.1  \\
2114.34 & \ion{Fe}{3}x2+\ion{Cr}{3} & $-$0.23  & \nodata & \nodata \\
2114.88 & \ion{Cr}{3}x2 & {\it +0.30}  & $-$0.3 	 & \nodata \\
2116.59 & \ion{Fe}{3}x2 &  fixed       & {\it 1.4}       & \nodata \\
2117.55 & \ion{Cr}{3}   & {\it +0.18}  & $-$0.4          & \nodata \\
2118.49 & \ion{Fe}{3}x2+\ion{Cr}{3} & {\it +0.15} & $-$0.7  & \nodata \\
2120.77 & \ion{Fe}{3}x4+\ion{Fe}{4} & $-$0.10  & \nodata & \nodata \\
2123.59 & \ion{Fe}{3}x3+\ion{Cr}{3} & $-$0.07  & $-$0.4  & \nodata \\
2124.16 & \ion{Fe}{3}x5+\ion{Co}{3} & {\it $-$0.47} & \nodata & \nodata \\
2125.18 & \ion{Fe}{3}x5+\ion{Fe}{4}+\ion{Mn}{3} & {\it $-$0.30} & $-$0.4 & \nodata \\ 
2126.14 & \ion{Mn}{3}   & $-$0.15      & \nodata & \nodata \\
2129.68 & \ion{Fe}{3}x3 & $-$0.15      & \nodata & \nodata \\
2134.83 & \ion{Fe}{3}x2 & {\it +0.40}  & $-$0.4  & \nodata \\
2136.36 & \ion{Fe}{3}x2 & {\it $-$0.60} & $-$0.9 & \nodata \\
\\
AVG	&		& $-$0.07  & $-$0.6	& $-$1.0  \\
1 $\sigma$	&	& 0.10     & 0.2  	& 0.3     \\
\enddata
\tablecomments{Results are relative to solar abundances in 
Grevesse \& Sauval (1998), e.g., 12+log(Fe/H)=7.50, 12+log(Mn/H)=5.53.  
Abundances are determined from spectrum syntheses using the parameters 
listed in Tables~\ref{atms} and \ref{abus}.  Dominant features 
only are identified.
Results that are $\ge$2\,$\sigma$ from the mean are
noted in italics and not included in the average. 
The five lines that have had their oscillator strengths adjusted based on the spectrum 
of HD\,36591 are listed as ``fixed'' 
and used {\it only} in the 
differential analyses (i.e., not included in the averages 
listed here). } 
\end{deluxetable}
\clearpage

\clearpage
\begin{deluxetable}{lrrrrr}
\tablecaption{ Boron Abundance Uncertainties \label{boron-unc}}
\tablewidth{0pt}
\tablehead{
\colhead{} & \colhead{AV\,304}    & \colhead{HD\,34078} & \colhead{HD\,36591} & 
\colhead{NGC\,346-637} \\[.2ex]                   
\colhead{} & \colhead{$\Delta$log($B$/H)} & \colhead{$\Delta$log($B$/H)} & 
\colhead{$\Delta$log($B$/H)} & \colhead{$\Delta$log($B$/H)}} 
\startdata
$\Delta$\teff\,= $\pm$750~K        & $\pm$0.1  & $\pm$0.2 & $\pm$0.03 & $\pm$0.2 \nl
$\Delta$\logg\,= $\pm$0.1          & $\pm$0.0  & $\mp$0.1 & $\pm$0.04 & $\pm$0.0 \nl 
$\Delta\xi_{\rm micro}=\pm$1~\kms & $\mp$0.0  & $\pm$0.0 & $\pm$0.00 & $\pm$0.0 \nl 
$\Delta\xi_{\rm macro}=\pm$2~\kms & $\pm$0.0  & $\pm$0.0 & $\pm$0.00 & $\pm$0.0 \nl 
Shift Continuum $\pm$2\%          & $\mp$0.1  & $\mp$0.2 & $\mp$0.12 & $\mp$0.2 \nl 
$^{11}$B = 2.5\,$^{10}$B          &   0.0     &   0.0    & $-$0.04   &  0.0     \nl 
$\Delta$V$_{\rm rad}=\pm$4~\kms   & $\pm$0.0  & $\pm$0.0 & $\pm$0.14 & $\pm$0.0 \nl  
($\Delta$V$_{\rm rad}$ \ion{Mn}{3}$\lambda$2065.9)\tablenotemark{^*} 
	& \nodata & ({\it $\pm$0.0}) & ({\it $\mp$0.09}) & \nodata \nl
\enddata
\tablecomments{$\Delta$log(B/H) was determined by setting \ion{Mn}{3}~$\lambda$2065.9 {\it a priori}
to the best fit (listed in italics in Table~\ref{abus}).
We expect the uncertainties in $v$\,sin\,$i$ to
be reflected in $\Delta\xi_{\rm macro}$.  For HD\,36591 and 
HD\,34078, the uncertainties in continuum placement 
and radial velocity are half of what is listed here; e.g. the 
uncertainty in HD\,36591 is 1\% and 2 \kms, yielding
$\Delta$log($B$/H) = 0.06 and 0.07, respectively. }  
\tablenotetext{^*}{Changing the radial velocity required a completely new fit,
i.e. letting \ion{B}{3} and \ion{Mn}{3} both vary.  
The new ``best'' \ion{Mn}{3} 2065.9\,\AA\ abundance
when radial velocity is varied
is listed here in italics for this parameter.}
\end{deluxetable}
\clearpage

\begin{deluxetable}{lcl}
\tablewidth{0pc}
\tablecaption{Present \& Local Galactic Boron Abundances \label {galb}}
\tablehead{
\colhead{Object} & \colhead{12+log(B/H)} & \colhead{Reference} }
\startdata
Solar & 2.8 &  Cunha \& Smith 1999 \\
---   &     & Zhai \& Shaw 1994 \\
---   &     & Anders \& Grevesse 1989 \\
Nebulae & $\geq$2.4 & Howk \etal 2000 \\
B-stars & 2.7  & Cunha \etal 1997 \\
--- & 2.4      & Proffitt \etal 1999 \\
--- & 2.3      & Proffitt \& Quigley 2001 \\
--- & 2.4      & Venn \etal 2001 \\
FG-stars & 2.6 & Cunha \etal 2000a \\  
---      & 2.5 & Cunha \etal 2000b \\  
\\
  & 2.5          & ADOPTED \\
\enddata
\end{deluxetable}
\clearpage


\clearpage
\clearpage

\clearpage
\clearpage

\clearpage
\clearpage

\clearpage
\clearpage

\clearpage
\clearpage

\clearpage
\clearpage

\end{document}